\documentclass[prl,twocolumn,showpacs]{revtex4}

\usepackage{graphicx,epsf,bm,bbm,color,amsmath,ulem}

	\newcommand{\beq}{\begin{equation}}
	\newcommand{\be}{\begin{equation}}
	\newcommand{\beqn}{\begin{eqnarray}}
	\newcommand{\eeq}{\end{equation}}
	\newcommand{\ee}{\end{equation}}
	\newcommand{\eeqn}{\end{eqnarray}}
	\newcommand{\nn}{\nonumber}
	\newcommand{\q}{{\bf q}}
	\newcommand{\ep}{{\epsilon}}
	\renewcommand{\k}{{\bf k}}
    \renewcommand{\r}{{\bf r}}

	\newcommand{\G}{{\bf G}}

\newcommand{\tb}[1]{\textbf{#1}}
\newcommand{\tr}[1]{\textrm{#1}}
\newcommand{\hb}{\hbar}

\newcommand{\f}{\frac}

\begin{document}

	
		\title{Bloch-Zener oscillations across a merging transition of Dirac points}
		\author{Lih-King Lim, Jean-No\"el Fuchs and Gilles Montambaux}
		\affiliation{Laboratoire de Physique des Solides, CNRS UMR 8502, Univ. Paris-Sud, F-91405 Orsay cedex, France.}

		\begin{abstract}
Bloch oscillations are a powerful tool to investigate spectra with Dirac points. By varying band parameters, Dirac points can be manipulated and merged at a topological transition towards a gapped phase. Under a constant force, a Fermi sea initially in the lower band performs Bloch oscillations and may Zener tunnel to the upper band mostly at the location of the Dirac points. The tunneling probability is computed from the low energy universal Hamiltonian describing the vicinity of the merging. The agreement with a recent experiment on cold atoms in an optical lattice is very good.
		\end{abstract}
	\maketitle

{\it Introduction.--}
Dirac points in energy bands occur in special two-dimensional (2D) condensed matter systems \cite{Dirac}, such as graphene \cite{CastroNeto}, nodal points in $d$-wave superconductors and surface states of three-dimensional (3D) topological insulators \cite{Hasan10}.
They are fascinating instances of ultra-relativistic behavior emerging as low-energy effective description of electrons in solids.
 Dirac points are band touching points that   carry a topological charge, namely a Berry phase $\pm \pi$. In most systems, Dirac points occur in dipolar pairs (the so-called fermion doubling).
Under variation of external parameters, it is possible to move these Dirac points and even make them merge. This merging signals a topological (Lifshitz) transition between a gapless phase with a disconnected Fermi surface to a gapped phase \cite{Hasegawa,Dietl,MontambauxUH,Wunsch}.
 For example, a uniaxial stress in   graphene   leads to a motion of the Dirac points but   the merging transition is not reachable \cite{CN}. The quasi-2D organic conductor $\alpha$-(BEDT-TTF)$_2$I$_3$ is a good candidate to observe this transition under pressure \cite{BEDT}, however it has not been realized yet.

Recently, a new type of experiment, realized with ultracold atoms loaded in a 2D optical lattice, has provided an alternative way to study Dirac points \cite{Esslinger}. By combining techniques of Bloch oscillations and adiabatic mapping of cold atoms, the band structure of the system can be studied with momentum resolution \cite{MorschOberthaler2006,Salger2007,Kling2011,Oelschlaeger2012}.
Specifically, the experiment of ETH Z\"urich \cite{Esslinger} utilizes such techniques for
a non-interacting Fermi gas in a tunable two-band system featuring Dirac points. Their existence is revealed through Landau-Zener (LZ) tunneling from the lower to the upper band. As the lattice amplitude is varied, a drastic change in the  transferred atomic fraction provides a qualitative signature of the Dirac points and their merging.

In this letter, we present a complete description of Landau-Zener tunneling through a pair of Dirac points, using a universal low-energy Hamiltonian describing the  merging transition \cite{MontambauxUH}.  We show how the transferred fraction provides a key signature of the merging transition and that it depends crucially on the direction of the motion with respect to the merging direction.
\begin{figure}[ht]
\begin{center}
\includegraphics[width=8cm]{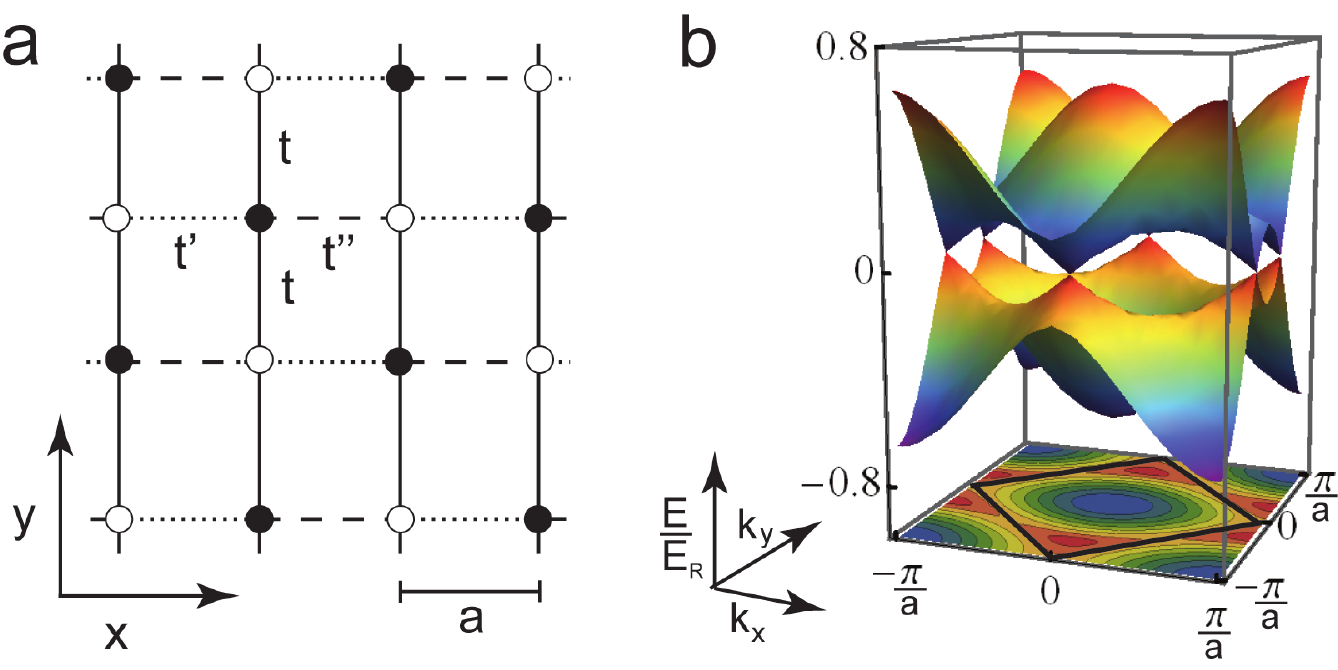}
\end{center}
 \caption{(color online). (a) Square lattice indicating the hopping amplitudes and the two inequivalent sites. (b) Band structure in the gapless D phase for $t'=t=0.2$, $t''=0.05$ in units of $E_R$, see text. The first Brillouin zone is indicated by the square.}
\label{fig:BZ}
\end{figure}
We find a very good agreement between the computed averaged LZ probabilities and the experimental data. Furthermore, new experimental signatures for varying Bloch oscillations and a coherent St\"{u}ckelberg interferometry are presented.

{\it Tight-binding model.--}
We consider a nearest-neighbor tight-binding model on a square lattice. The four hopping amplitudes between neighbors are taken as $t$, $t$ along $y$- and $t'$, $t''$ along $x$-direction (see Fig. \ref{fig:BZ}). When $t'\neq t''$, there are two inequivalent sites -- called $A$ and $B$ -- per unit cell giving rise to two bands. When $t''=0$, a link is broken realizing a brick-wall lattice, which has the same connectivity as the honeycomb lattice albeit with a rectangular geometry. When $t'=t''$, it is a standard square lattice with anisotropic amplitudes along $x$ and $y$ and a single site per unit cell. The Hamiltonian (with nearest neighbor distance $a\equiv 1$ and $\hbar \equiv 1$) reads
\be H= \left(
  \begin{array}{cc}
    0 & f(\k)  \\
     f^*(\k)  & 0 \\
  \end{array}
\right)\label{Hdek} \ee
with $f(\k)=-(t e^{i k_y}+ t e^{-i k_y}+t'e^{ik_x}+t''e^{-ik_x})$ where the hopping amplitudes are positive and $\k=(k_x,k_y)$ is the Bloch wavevector.
The energy spectrum is given by $\ep(\k)= \pm |f(\k)|$. It features two Dirac cones when $t'+t''<2t$ (gapless D phase) and a gap when $t'+t''>2t$ (gapped G phase). At $t'+t''=2t$ the Dirac points merge at $\k=(0,\pi)$ and there is a single touching point between the two bands.
When $t'=t''$, the band structure is that of a square lattice, with lines of Dirac points (L phase), which becomes isotropic when $t'=t''=t$ (I phase). This model can therefore describe the transition between the G and D phase, and moreover, the crossover from the D to L phase. In the following, energies are measured in units of the recoil energy $E_R= \pi^2\hbar^2/(2ma^2)$ where $m$ is the atomic mass.

{\it Mapping to the universal Hamiltonian.--}
In a crystal which is time-reversal and inversion symmetric,
 merging can only occur at $\G/2$ points  where $\G$ is a reciprocal lattice vector. Near such a point, it is possible to write a minimal
low energy Hamiltonian that captures the topological transition and describes both Dirac cones at once \cite{MontambauxUH}. Close to the merging, an expansion   for small $\q=\k-\G/2$ gives rise to an effective Hamiltonian which has the universal form
\be {\cal H} = \left(
                 \begin{array}{cc}
                   0 & \Delta_* + {q_y^2 \over 2 m^*} - i c_x q_x  \\
                   \Delta_* + {q_y^2 \over 2 m^*} + i c_x q_x & 0 \\
                 \end{array}
               \right) \label{HU}\ee
and a spectrum $\ep= \pm \sqrt{({q_y^2 \over 2 m^*}+ \Delta_*)^2 + c_x^2 q_x^2} $. The model depends on three independent parameters $\Delta_*$, $m^*$ and $c_x$. The merging transition is driven by the parameter $\Delta_*$ hereafter called the merging gap. When $\Delta_*<0$, the spectrum contains two Dirac points at $\q_D=(0,\pm \sqrt{2m^* |\Delta_*|})$ and $|\Delta_*|$ represents the energy of the saddle points connecting them, which is located at $\q=0$  (D phase). When increasing $\Delta_*$ towards $0$,   the two Dirac points approach each other along the $q_y$ direction until they merge when $\Delta_*=0$. Exactly at the merging, there is a single touching point between the two bands,  with a semi-Dirac (or hybrid) spectrum $\ep=\pm\sqrt{(q_y^2/2m^*)^2+(c_x q_x)^2}$ \cite{Dietl,Pickett}. By increasing the driving parameter still further, a true gap of magnitude $2\Delta_*>0$ opens at $\q=0$ (G phase).
We map  the tight-binding to the universal model by comparing their energy expansions near the Dirac points \cite{MontambauxUH}. In the D phase, we find   $\Delta_*= t'+t'' - 2 t$  and the Dirac cone velocities are $c_x= t'-t'' $ and $c_y = \sqrt{4 t^2 - (t'+t'')^2}$, so that the mass is obtained from $m^*= -2 \Delta_*/ c_y^2=2/(2t+t'+t'')$. In the G phase, $\Delta_*$ and $c_x$ are unchanged and $m^*=1/(2 t)$.

\begin{figure}[h!]
\begin{center}
\includegraphics[width=6.5cm]{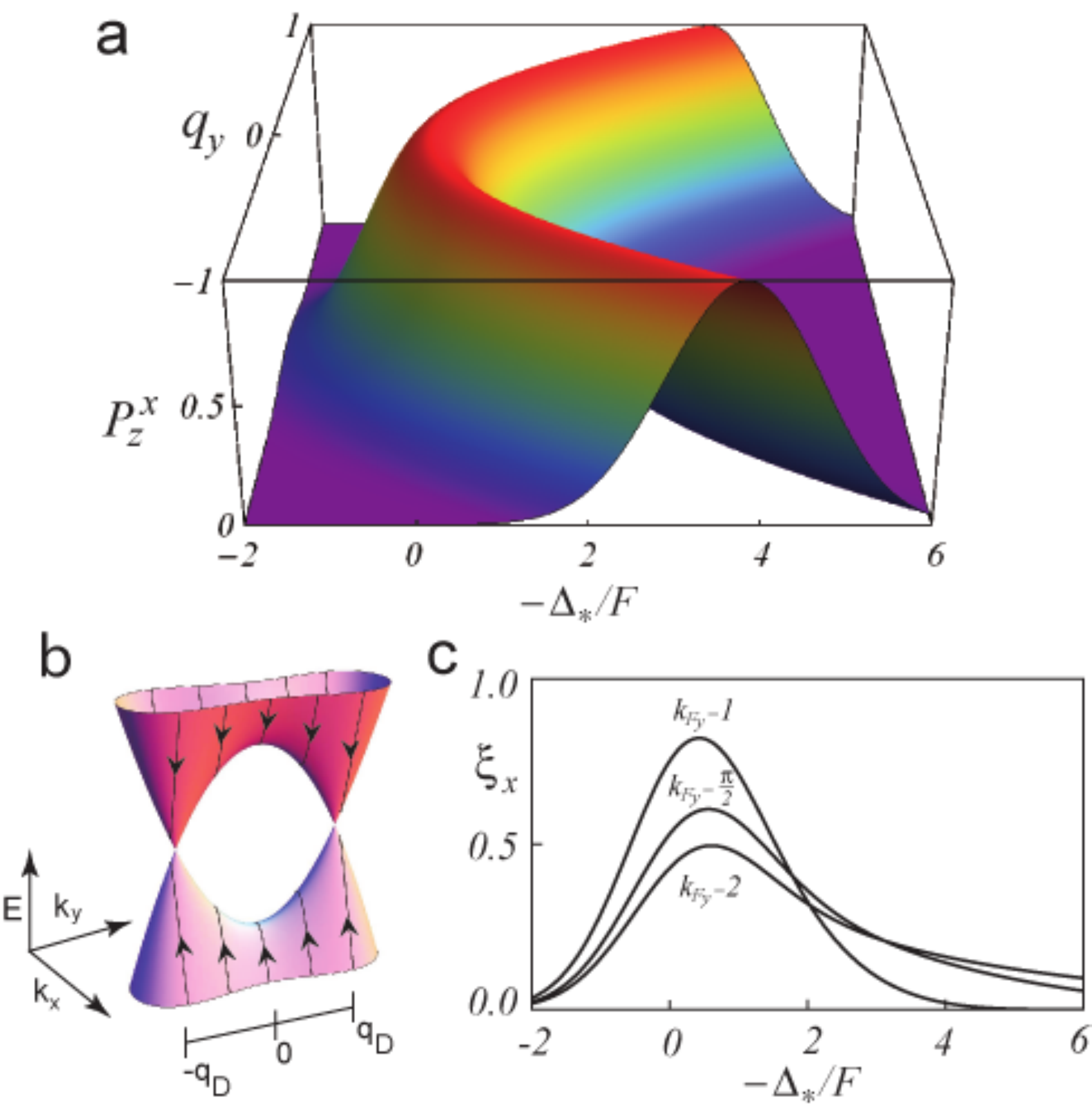}
\end{center}
\caption{(color online). {\it Motion along  $k_x$}. (a) LZ probability $P_Z^x$ (eq. (\ref{pzx})) as a function of   $-\Delta_*/F$ and of the transverse momentum $q_y$. Here $c_x/F\approx 0.5$ and $m^* F\approx 0.13$. (b) Trajectories along the $k_x$ direction. (c) Transferred fraction to the upper band $\xi_x$ as a function of  $-\Delta_*/F$ for different sizes $k_{Fy}$ of the cloud.  }
 \label{fig:zener-kx}
\end{figure}
{\it  Landau-Zener tunneling with Dirac cones.--} Consider atoms initially in the lower band performing Bloch oscillations  under the influence of a constant applied force $F$. By accelerating these atoms
in the vicinity of a Dirac point, their tunneling probability to the upper band is finite, a problem considered by Landau and Zener \cite{Zener}. In the following, the universal low-energy Hamiltonian   is used to compute the interband transition probability.

{\it Motion along the $k_x$ direction: single Dirac cone.--}
In the D phase, an atom moving along the $k_x$ direction encounters at most one Dirac cone during a single Bloch oscillation (Fig. \ref{fig:zener-kx}b).
The LZ probability for such a linear avoided band crossing is given by \cite{Zener}
\be P_Z^x=  e^{ \displaystyle -\pi  {(\textrm{gap/2})^2 \over  c_x F}}=e^{ \displaystyle -\pi  {({q_y^2 \over 2 m^*} + \Delta_* )^2 \over  c_x F}}\label{pzx} \ee
where $q_y$ is   the position with respect to the merging point $\G/2=(0,\pi)$.  Note that $P_Z^x=1$ for $q_y =\pm q_D= \pm \sqrt{-2 m^* \Delta_*}$, positions of the two Dirac points. Actually Eq. (\ref{pzx}) is not only valid in the D but extends to the G phase across the merging transition. This quantity is shown in Fig. \ref{fig:zener-kx}a as a function of the transverse momentum $q_y$ and $\Delta_*$.

As the experiment is  performed with a cloud of non-interacting fermions,
we need also to average the LZ probability over the initial distribution of atoms. We consider a 2D cloud of harmonically trapped fermions at zero temperature for   a filling fraction sufficiently smaller than half-filling.
The energy spectrum close to $\k=0$ is $\ep(\k)\approx k_x^2/(2m_x)+k_y^2/(2m_y)$ (as measured from the band bottom) with the band masses $m_x=(2t+t'+t'')/[4t't''+2t(t'+t'')]$ and $m_y=1/(2t)$. The semiclassical energy of an atom is therefore $\ep(\k,\r)=k_x^2/(2m_x)+k_y^2/(2m_y)+(m_x \omega_x x^2 + m_y \omega_y^2 y^2)/2$ where $\omega_x/2\pi$ and $\omega_y/2\pi$ are the trapping frequencies \cite{trappingfrequency}. The fraction $\xi_x$ of atoms transferred to the upper band is then given by the averaged probability $\xi_x=\langle P_Z^x \rangle$ where
\be \langle \cdots \rangle = {\int_{\ep(\k,\r) < \ep_F} d k_x d k_y dx dy   \cdots  \over  \int_{\ep(\k,\r) < \ep_F} d k_x d k_y dx dy    } \label{average} \ee
and $\ep_F=k_{Fx}^2/(2m_x)=k_{Fy}^2/(2m_y)$ is the Fermi energy, which defines $k_{Fx}$ and $k_{Fy}$ \cite{supplemental}.
\begin{figure}[h!]
\begin{center}
\includegraphics[width=6.5cm]{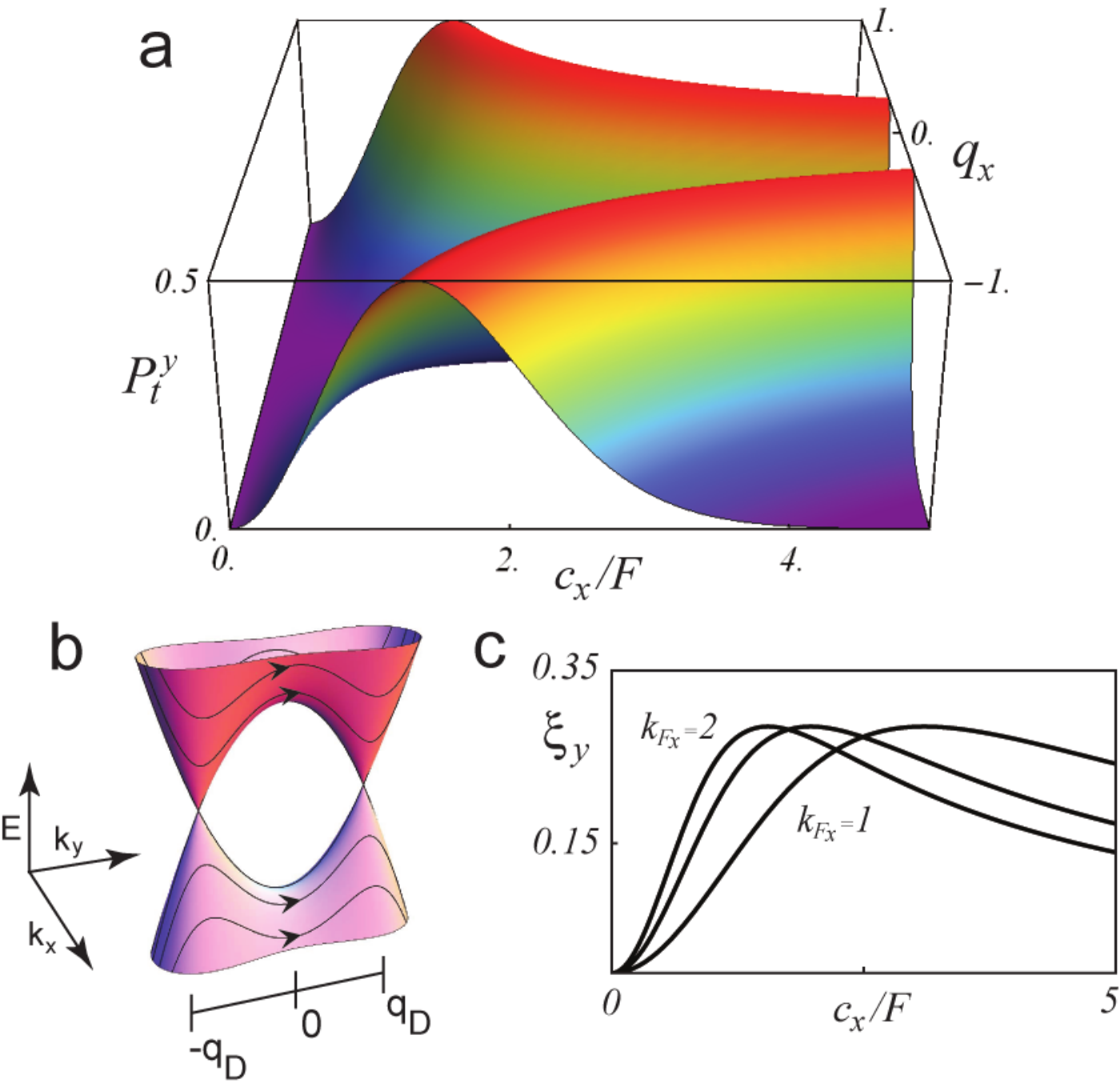}
\end{center}
\caption{(color online). {\it Motion along $k_y$.} (a) Total probability $P_t^y$ for atoms tunneling to the upper band (eq. (\ref{Pty})) as a function of  $c_x/F$ and of the transverse momentum $q_x$. Here $\Delta_*/F=-5$ and $m^*F\approx 0.13$. (b) Double LZ tunneling along the $k_y$ direction. (c) Transferred fraction  $\xi_y$ as a function of $c_x/F$ for various sizes $k_{Fx}$ of the initial cloud. }
 \label{fig:zener-ky}
\end{figure}
The transferred fraction $\xi_x$ as a function of  $\Delta_*$ and of the size $k_{Fy}$ of the cloud is shown in Fig. \ref{fig:zener-kx}c. For a cloud of finite size $k_{Fy}$, only a finite proportion of atoms may tunnel to the upper band when $\Delta_*<0$.

{\it Motion along the $k_y$ direction: double Dirac cone.--}
In the G phase, the tunneling probability is vanishingly small. In the following, we concentrate on the D phase where atoms performing one Bloch oscillation  in the $k_y$ direction have the possibility to encounter two   inequivalent  Dirac cones successively. The scenario  is therefore richer than before since the tunneling process implies two successive Landau-Zener events (see Fig. \ref{fig:zener-ky}b). The probability $P_Z^y$ associated with each LZ event is now
\be P_Z^y \equiv e^{- 2 \pi \delta}= e^{\displaystyle -   \pi {c_x^2 q_x^2  \over c_y F}}= e^{\displaystyle -   \pi {c_x^2 q_x^2  \over  F\sqrt{2|\Delta_*|/m^*}}} \label{PZ2} \ee
which defines the adiabaticity parameter $\delta$. In the following, we calculate the total interband probability $P_t^y$ associated with the two successive events, in the limit where they can be considered independent.
Quantitatively, the LZ tunneling time $\sim \textrm{max}(\sqrt{\delta},\delta)/c_x q_x$ \cite{revue} should be shorter than the time $2\sqrt{2m^*|\Delta_*|}/F$ it takes an atom to travel between the two Dirac points, i.e. not too close to the merging transition.

First assuming that the two tunneling events are incoherent, we combine the probabilities to find the interband transition probability \cite{supplemental, revue}
\be P_t^y = 2 P_Z^y (1 - P_Z^y) \label{Pty} \ee
which is shown in Fig. \ref{fig:zener-ky}a as a function of $c_x$ and the transverse momentum $q_x$.
Notice that $P_t^y$ vanishes when $q_x=0$ because $P_Z^y=1$. For an initial cloud of size $k_{Fx}$, the transferred fraction is $\xi_y=\langle P_t^y \rangle$ where the average is defined in Eq. (\ref{average}) \cite{supplemental}.
The result is shown in Fig. \ref{fig:zener-ky}c.

\begin{figure}[h!]
\begin{center}
\includegraphics[width=8.5cm]{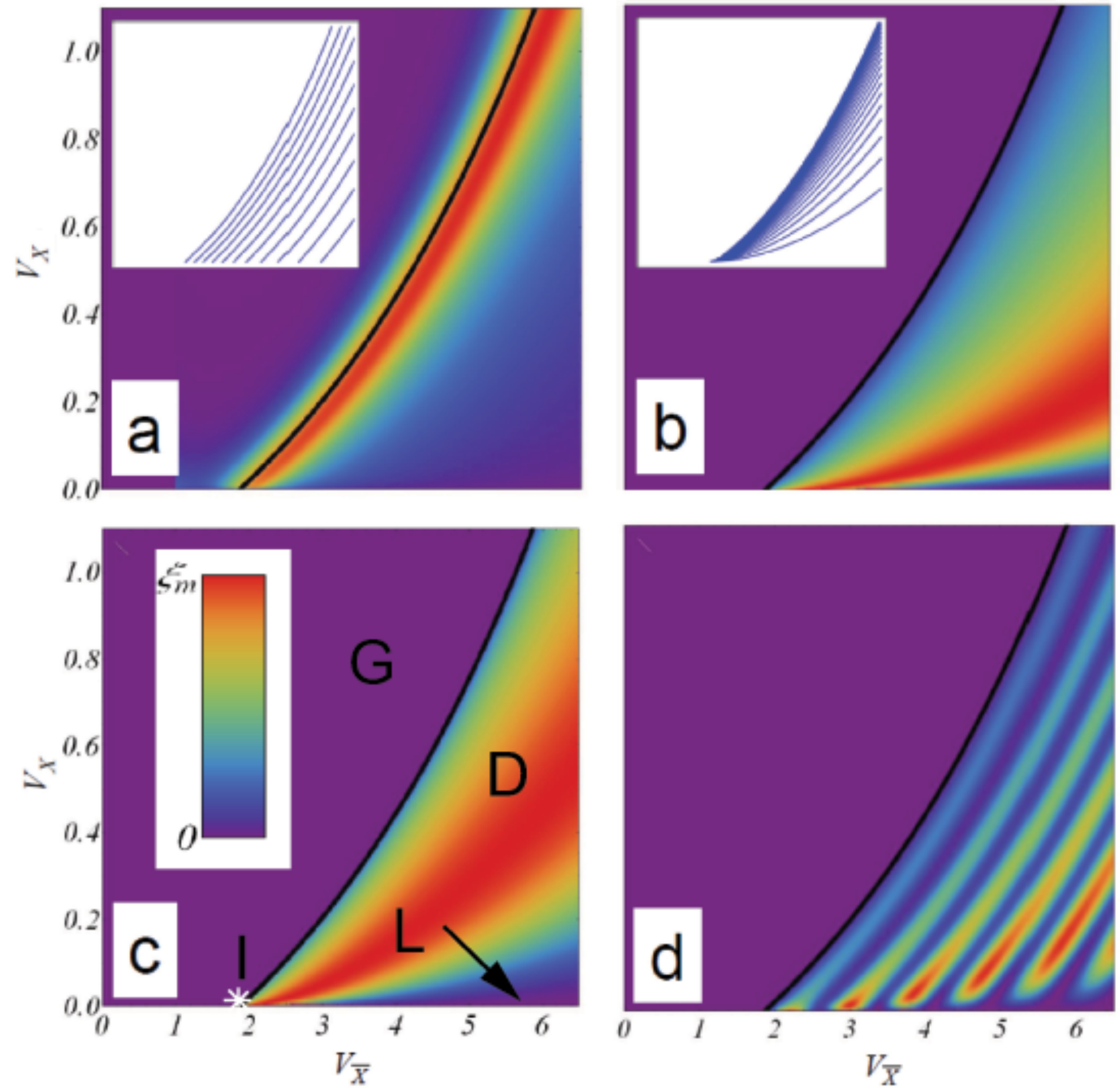}
\includegraphics[width=6cm]{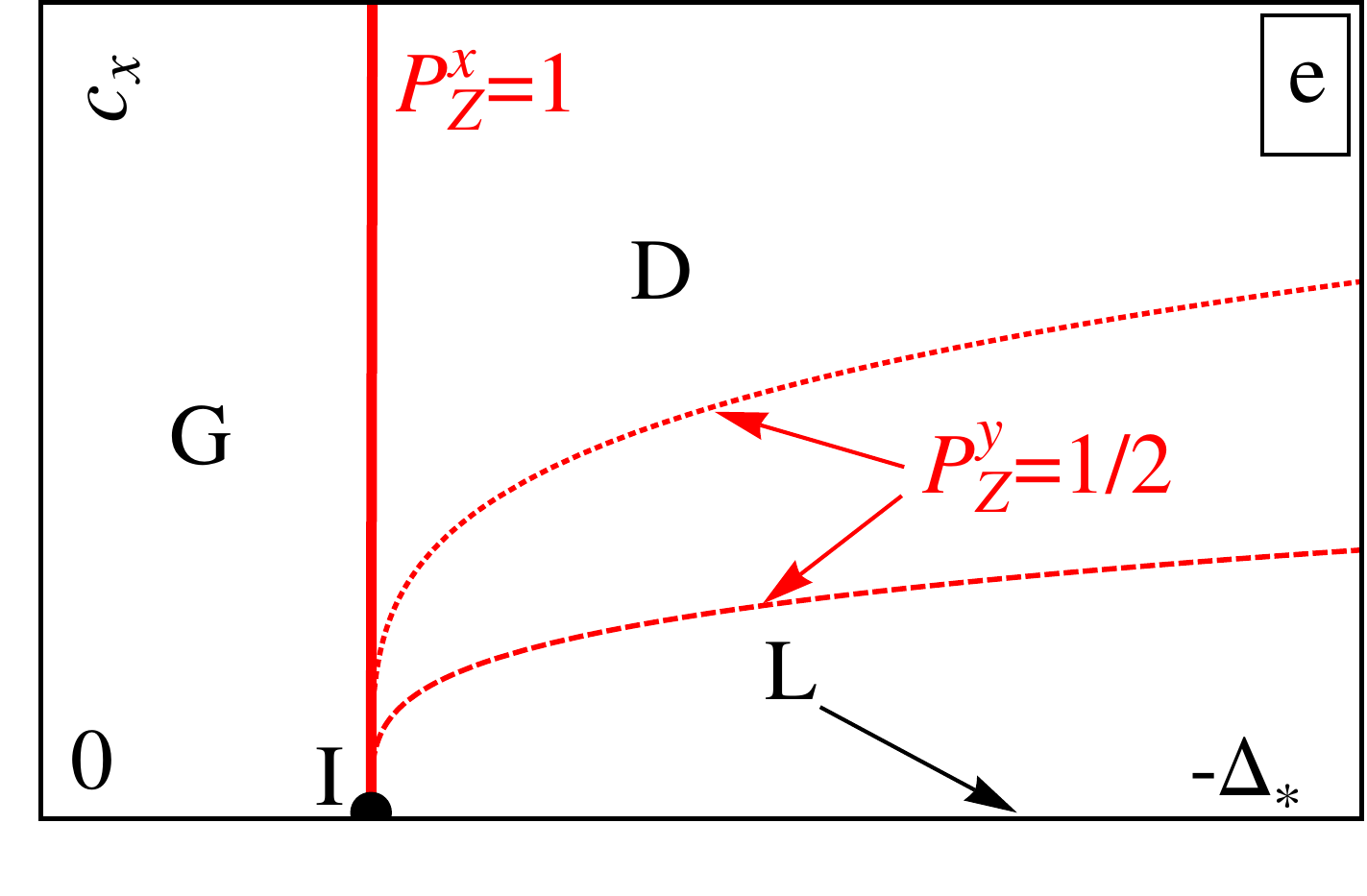}
\end{center}
\caption{(color online). a) Transferred fraction $\xi_x$  as a function of the optical lattice parameters $V_{\bar{X}}$ and $V_X$     (here $k_{Fy}=\pi/2, F=0.02$). Inset: lines of constant $\Delta_*$. b) Transferred fraction $\xi_y$ ($k_{Fx}=2, F=0.02$). Inset: lines of constant $c_x^2/c_y$. c) Same as (b) with $F=0.1$.  d) Same parameters as (b) taking coherence into account and leading to St\"uckelberg oscillations.  e) Phase diagram in the $(-\Delta_*,c_x)$ plane showing the G, D, L, and I phases (see text).   $P_Z^x=1$ along the merging transition (continuous line). The crossover line $P_z^y=1/2$ corresponds to $c_x\propto F^{1/2}|\Delta_*|^{1/4}$ and is plotted for two different forces (dashed and dotted lines). The color code for (a)-(d) is such that $\xi_m=0.5,0.3,0.3,$ and $0.6$, respectively. The black line in (a)-(d) indicates the merging transition $\Delta_*=0$.}
 \label{fig:manip}
\end{figure}

{\it Comparison to the experiment.--} The ETH experiment is performed on a harmonically trapped 3D Fermi gas loaded in a 2D optical trap \cite{Esslinger}. Our model treats a trapped 2D Fermi gas in a 2D band structure. The optical lattice potential is $V(x,y)=-V_{\bar{X}}\cos^2(\pi x+\theta/2)-V_X \cos^2 (\pi x) - V_Y\cos^2(\pi y) -2\alpha \sqrt{V_X V_Y}\cos(\pi x) \cos(\pi y)$, where $\alpha=0.9$, $\theta=\pi$, and the laser wavelength is $2a$ with amplitudes $V_Y=1.8$, $0\leq V_{\bar{X}}\leq 6.5$ and $0\leq V_X\leq 1$. To make a precise comparison, we perform single-particle numerical band structure calculation provided by the 2D optical potential using a truncated plane-wave expansion, and establish a map between the optical lattice parameters and that of the universal Hamiltonian of Eq. (\ref{HU}) for $\Delta_*<0$. The latter is done by fitting the parameters of the universal Hamiltonian to the two lowest lying energy bands in the vicinity of the Dirac cones \cite{supplemental}. The corresponding $(t,t',t'')$-tight-binding model is then obtained by the analytical mapping (see paragraph after Eq. (\ref{HU})) -- the main features that are not captured are the Dirac cones tilting and the particle-hole asymmetry, which appear not to be relevant. Qualitatively, $t$ stays roughly constant in the considered range of optical lattice parameters. The hopping $t'$ increases when $V_X-V_{\bar{X}}$ increases, whereas $t''$ increases when $V_X+V_{\bar{X}}$ decreases.
Finally we compute the transferred fractions $\xi_x$ and $\xi_y$ as a function of $V_X$ and $V_{\bar{X}}$, shown in Fig. \ref{fig:manip}. For the experimental parameters $F=0.02$, $k_{Fy}\simeq \pi/2$ and $k_{Fx}\simeq 2$ (corresponding to $\ep_F\approx 0.4$), we find a striking agreement with Figs. 4a and 4b of Ref. \cite{Esslinger}.

{\it Discussion.--} First consider the case of the motion along $k_x$. The line of maximum transfer probability $\xi_x$  (red region in Fig. \ref{fig:manip}a) corresponds to a maximal LZ probability $P_Z^x \approx 1$ for a large number of atoms. Playing with the averaging order gives  $\xi_x \approx \exp[-\pi \langle (q_y^2/2m^*+\Delta_*)^2\rangle/(c_x F)]$, in which $\langle q_y^2\rangle=k_{Fy}^2/6$ and $\langle q_y^4\rangle=k_{Fy}^4/16$. Its maximum occurs when $\Delta_*  \simeq - \langle q_y^2 \rangle/(2 m^*)$, which explains why it is near the merging line $\Delta_*= 0$, but slightly inside the D phase, as seen experimentally. The transferred fraction $\xi_x$ is a symmetric function of its natural variable $\Delta_*$ -- when doing the average properly it is actually slightly asymmetric -- and its width reduces when decreasing $c_x$ (by decreasing $V_X$). Both features are seen experimentally, see Fig. 4a of Ref. \cite{Esslinger}. Obviously, the width and the maximum should increase when increasing $F$.

Adding a staggered on-site  potential $\pm \Delta/2$  opens a gap $\Delta$ in the spectrum. Experimentally, this gap  is controlled by the parameter $\theta$ of the  potential $V(x,y)$. It was found that the decay of the transferred fraction $\xi_x$ as a function of $\theta$ is well fitted by a gaussian (see Fig. 2b of Ref. \cite{Esslinger}). Here, we prove that it is of the form $\xi_x(\Delta)=\xi_x(0)\exp(-\pi  \Delta^2/(4 c_x F))$ and that $\Delta\approx 4.3 (\theta/\pi-1)$ for the experimental parameters $V_{\bar{X},X,Y}=[3.6,0.28,1.8]$.

Next, consider the case of the motion along $k_y$. The interband transition probability (Eq. \ref{Pty}) is a non monotonous function of the LZ probability $P_Z^y$ and it is  maximum when $P_Z^y=1/2$.  This explains the  existence of the maximum of $\xi_y$ well {\it inside} the D phase (red region in Fig. \ref{fig:manip}b). When playing with the averaging order, $\xi_y\approx e^{-X} (1 - e^{-X})/2$ where $X\equiv \pi c_x^2 \langle q_x^2 \rangle /c_y F$. The maximum occurs when $e^{-X} \approx 1/2$ i.e. when $c_x^2/c_y \simeq F \ln 2 /(\pi \langle q_x^2\rangle)  = \textrm{const}$. Compared to the previous case, $\xi_y$ is a much more asymmetric function of its natural variable $X\propto c_x^2/c_y$: the decay at large $X$ is slower than at small $X$  (the signal extends more towards large $c_x$, which is seen experimentally). In contrast to the previous case, the position of the line of maxima depends on $F$, as shown in Figs.  \ref{fig:manip}b and 4c, but its amplitude is almost independent. Its width decreases when $c_x \rightarrow 0$. Furthermore, $\xi_y$ vanishes as $V_X\rightarrow 0$, eventually reaching the square lattice limit (L phase).

Since the two LZ events along $k_y$ are expected to be coherent, St\"uckelberg interferences in the transferred fraction $\xi_y$ should be observable. Eq. (6)   should indeed be replaced by $P_t^y=4P_Z^y(1-P_Z^y)\cos^2(\varphi/2+\varphi_d)$ where $\varphi=4\int_0^{q_D}  dq_y \sqrt{(\Delta_*+q_y^2/2m^*)^2+c_x^2q_x^2} /F$ is a dynamically acquired phase in between  the two tunneling events and $\varphi_d=-\pi/4+\delta (\ln \delta -1) + \textrm{arg}\,\Gamma (1-i\delta)$ is a phase delay given in terms of the gamma function and $\delta$ is defined in Eq. (5)  \cite{revue,supplemental}. Averaging $P_t^y$ over the 2D atomic distribution gives the transferred fraction $\xi_y$ shown in Fig.  \ref{fig:manip}d. However, such interferences are not observed in the ETH experiment, which we attribute to averaging over the third spatial direction.
Briefly, interference fringes in Fig. \ref{fig:manip}d are lines of constant $\Delta_*$ with a fringe spacing $\sim 0.04 E_R$. For the experimentally given trapping frequency in the $z$ direction, we estimate that $\Delta_*$ varies along $z$ by $\sim 0.03E_R$. This should be enough to wash out the interferences.

{\it Conclusion.--}
Landau-Zener tunneling conveniently probes the energy spectrum in the vicinity of Dirac points. Depending on the direction of the applied force, the atoms experience a LZ transition through a single or a pair of Dirac cones. We calculated the transferred fraction in the framework of the universal Hamiltonian describing the merging transition, and found a very good agreement with the ETH experiment. To summarize, the important parameters are the merging gap $\Delta_*$ and the velocity $c_x$  perpendicular to the merging direction. A simplified phase diagram is shown in Fig. \ref{fig:manip}e. Although the transfer through a single Dirac cone probes the merging transition, the transfer through a pair of cones signals a double Landau-Zener event inside the D phase and a cross-over towards the L phase.

As perspectives, we expect St\"uckelberg interferences to be observable in the strictly 2D regime. Furthermore, it should now be possible to tune the optical lattice right at the merging transition and to study the semi-Dirac spectrum, thus opening the way to explore new phenomena. For example, applying an artificial $U(1)$ gauge potential \cite{dalibard11, bloch11} should reveal unusual Landau levels \cite{Dietl}.

\acknowledgments{We acknowledge support from the Nanosim Graphene project under grant number ANR-09-NANO-016-01}.

\newpage
\section*{Appendix}

\bigskip

\renewcommand{\thefigure}{A\arabic{figure}}
 \setcounter{figure}{0}
\renewcommand{\theequation}{A.\arabic{equation}}
 \setcounter{equation}{0}
 \renewcommand{\thesection}{A.\Roman{section}}
\setcounter{section}{0}

\section{Numerical band structure calculation}
In the experiment \cite{EsslingerA}, a 2D tunable optical potential of the form
\beqn
V(x,y)&=&-V_{\bar{X}} \cos^2 (k x+\theta/2)-V_{X} \cos^2(k x)\nn\\
&&-V_{Y} \cos^2(k y)-2\alpha \sqrt{V_{X}V_{Y}} \cos(k x)\nn\\
&&\times \cos(k y) \cos(\varphi)
\eeqn
is utilized. Here, $V_{\bar{X}}$, $V_{X}$ and $V_{Y}$ are proportional to the tunable laser intensities, $\alpha=0.9$ is the visibility of the interference pattern, and other parameters $\theta=\pi, \varphi=0$ are fixed for the most part in the experiment. $k=2\pi/\lambda$ is the laser wave vector, and the nearest neighbor lattice distance used in the main text $a=\lambda/2$. The Bravais lattice of the optical potential is given by
$\tb{a}_1=(\lambda/2)\, (1,-1),\tr{\ }\tb{a}_2=(\lambda/2)\,(1,1),$ and the corresponding reciprocal vector is given as
$\tb{b}_1=(2\pi/\lambda)\,(1, -1),\tr{\ }\tb{b}_2=(2\pi/\lambda)\,(1,1)$. The single-particle bandstructure is obtained by numerically solving the 2D Schr\"{o}dinger equation
\beqn
\biggl[-\f{\hb^2}{2 m_0}\Delta+V(\tb{r})\biggr]\psi (\tb{r})=E\psi(\tb{r}).
\eeqn
By writing the wavefunction in Bloch's form $\psi_{\tb{k}}(\tb{r})=\exp(i\tb{k}\cdot\tb{r})\sum_{m,n\in \tr{Z}}\,u(m,n)\,\exp(i m\tb{b}_1\cdot\tb{r}+i n \tb{b}_2\cdot\tb{r})$,
the Schr\"{o}dinger equation in terms of the Fourier components $u(m,n)$ becomes
\beqn
&&\bigl\{\f{\hb^2}{2 m_0}\bigl[ (k_x+2\pi (m+n)/\lambda)^2+(k_y+2\pi (m-n)/\lambda)^2 \bigr]\nn\\
&&\,\,\,+\delta_1\bigr\}u(m,n)+\bigl\{ \delta_2\,u(m-1,n-1)+\delta_3\,u(m-1,n)\nn\\
&&\,\,\,+\delta_4\,u(m-1,n+1)+\delta_3\,u(m,n-1)+\delta_3\,u(m,n+1)\nn\\
&&\,\,\,+\delta_4\,u(m+1,n-1)+\delta_3\,u(m+1,n)\nn\\&&\,\,\,+\delta_2\,u(m+1,n+1)  \bigr\}=E\,u(m,n),
\eeqn
where the various constants are defined as $\delta_1=-(V_{\bar{X}}+V_{X}-V_{Y})/2$, $\delta_2=(V_{\bar{X}}-V_X)/4$, $\delta_3=-0.45 \sqrt{V_{X} V_{Y}}$ and $\delta_4= V_{Y}/4$. We truncate the Fourier coefficients $u(m,n)$ at $|m|,|n|=2$ and solve the resulting $25$ secular equations in \textit{Mathematica} to obtain the discrete energy bands $E_i(\tb{k})$ for each $\tb{k}$. A convergence test has been carried out for the numerical bandstructure calculation up to a truncation at $|m|,|n|=3$, which yields an excellent agreement with the $|m|,|n|=2$ case.

\begin{figure}
\begin{center}
\includegraphics[width=5.cm]{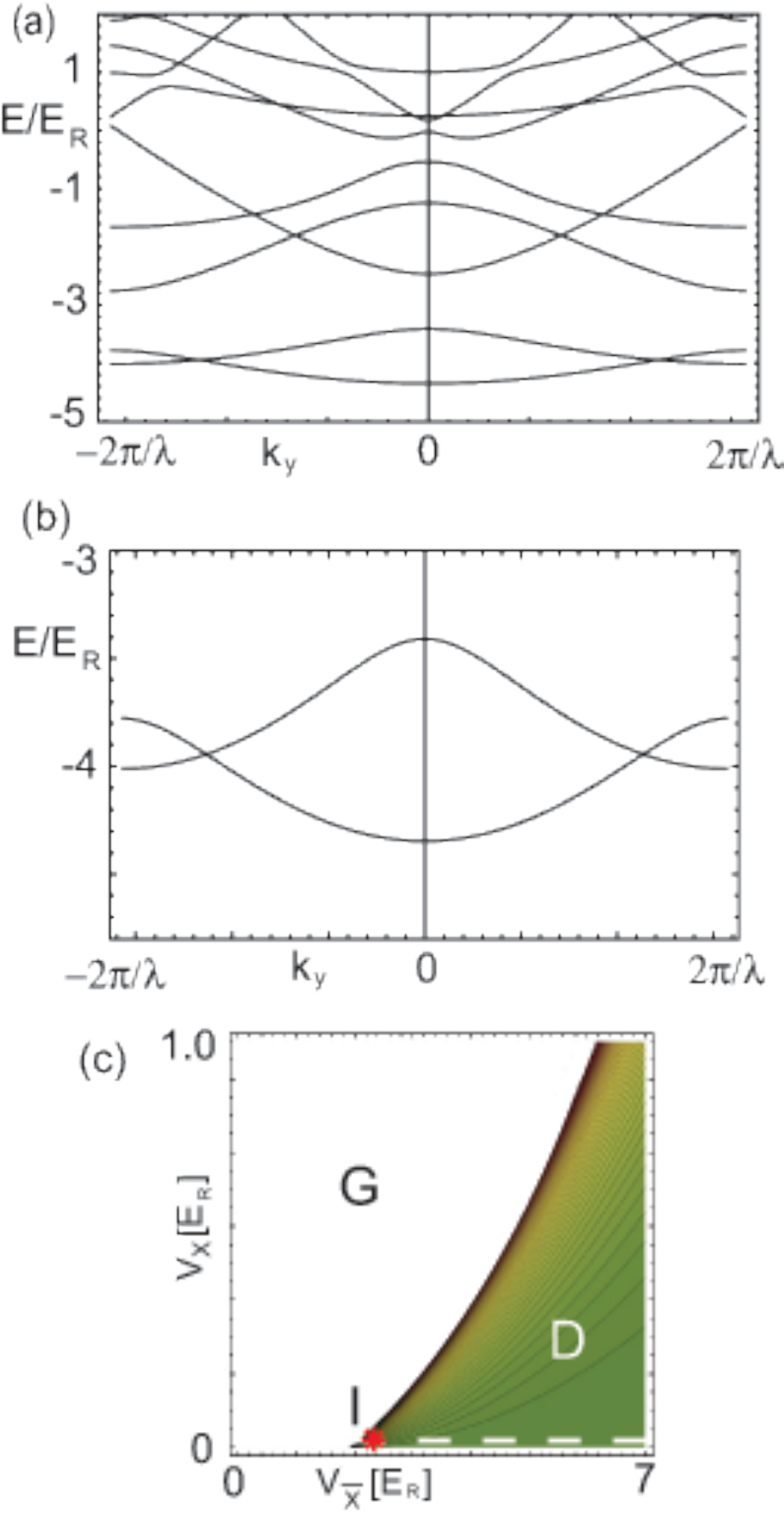}
\end{center}
\caption{(a) Bandstructure for $V_{\bar{X}}= 5 \, E_R$, $V_X=0.3 \, E_R$, $V_Y=1.8\, E_R$ in the $k_y$-direction with $k_x=0$. (b) The two lowest bands featuring two inequivalent Dirac points. (c) The three phases $D,G$ and $L$ (dashed line) realized with the optical potential. The starred point indicates the position where the isotropic square lattice model (I phase) is realized. }
\end{figure}

A typical bandstructure for a particular set of laser amplitudes is shown in Fig. A1(a), in the convenient recoil energy unit $E_R=h^2/2m_0\lambda^2$ where $m_0$ is the atomic mass. From the two lowest bands, see Fig. A1(b), we then determine the magnitude of the gap $\Delta_*$, the distance between the two Dirac points $q_D$ and the slope $c_x$ around the Dirac points in the $x$ direction as a function of $V_{\bar{X}}$ and $V_X$, for a fixed $V_Y=1.8\, E_R$. The interpolation formulae in the gapless phase are obtained as follow:
\beqn
&&|\Delta_* (V_{\bar{X}},V_X)| = \bigl( 0.1134+0.0218\, V_X-0.0205\, V_X^2 \bigr)\nn\\&&\times \ln \bigl[1+\bigl(V_{\bar{X}} -3.8234\, \ln [2.7560\, V_X+1.6325] \bigr)^{1.2}   \bigr],\nn\\
&&q_D (V_{\bar{X}},V_X)=\bigl(0.65+0.1 \,V_X\bigr)\nn\\&&\times \bigl( V_{\bar{X}} -3.8234\, \ln [2.7560\, V_X+1.6325]\bigr)^{0.42+0.05\, V_X},\nn\\
&&c_x(V_{\bar{X}},V_X)=1.1204\, e^{-0.3023 \, V_{\bar{X}}}\nn\\&&\times V_X^{\,\,0.3791 \ln [1.8380 \, V_{\bar{X}}-0.7251]}.
\eeqn
The slope in the $y$-direction around a Dirac cone is then given by $c_y=2 |\Delta_*| /q_D$. Furthermore, the boundary between the gapped and the gapless phase is given by
\beq
V_{\bar{X}}=a_1 \, \ln \bigl[ a_2\, V_X+a_3 \bigr]
\eeq
with $a_1=3.8234$, $a_2=2.7360$, $a_3=1.6325$. After translating into the more intuitive anisotropic square lattice model with $t,t',t''$ hopping amplitudes (see the paragraph after Eqn (2) in the main text), we then locate the various phases that are realized as a function of the laser amplitudes, see Fig. A1(c).

\section{Averaging over the atomic distribution}
\subsection{Motion along the $k_x$ direction: single Dirac cone}
As the experiment is done with a cloud of non-interacting Fermi gas,
we need to average the Landau-Zener probability over the distribution of atoms. Taking a 2D cloud of harmonically trapped atoms
at zero temperature filled up to the Fermi energy $\ep_F$ (measured from the band bottom),
the fraction of atoms transferred to the upper band is given by the averaged tunneling probability $\langle P_Z^x \rangle$ where the average is defined as
\be \langle \cdots \rangle = {\int_{\ep(\k,\r) < \ep_F} d k_x d k_y dx dy   \cdots  \over  \int_{\ep(\k,\r) < \ep_F} d k_x d k_y dx dy    } \label{average} \ee
Since the low energy expansion of the spectrum $\ep(\k)$ associated with the tight-binding Hamiltonian (1) is $\ep(\k)= t t'k_x^2/(2 t+t')+ t k_y^2$ (as measured from the band bottom), the semiclassical energy of an atom in the 2D anisotropic harmonic oscillator is
$\ep(\k,\r)=\frac{k_x^2}{2m_x}+ \frac{k_y^2}{2m_y}+\frac{1}{2}(m_x \omega_x x^2 + m_y \omega_y^2 y^2)$ , where $m_x=(2 t+t')/(2t t')$ and $m_y=1/(2t)$ are the band masses and $\omega_x/2\pi$ and $\omega_y/2\pi$ the trapping frequencies \cite{trappingfrequencyA}, the average is given by
\be \langle \cdots \rangle = {16 \over 3\pi k_{Fy}^4} \int_0^{k_{Fy}}  d k_y   (k_{Fy}^2 - k_y^2)^{3/2} \ldots   \label{moyenne} \ee
where $\ep_F\equiv \frac{k_{Fy}^2}{2m_y}=\frac{k_x^2}{2m_x}+ \frac{k_y^2}{2m_y}+\frac{1}{2}(m_x \omega_x x^2 + m_y \omega_y^2 y^2)$ defines the 2D Fermi surface.

\subsection{Motion along the $k_y$ direction: double Dirac cone}
For an initial cloud of size $k_{Fx}$, the transferred fraction is $\langle P_t^y \rangle$ where the average is defined in (\ref{average}) and becomes
\be \langle \cdots \rangle = {16 \over 3\pi k_{Fx}^4}\int_0^{k_{Fx}}  d k_x   (k_{Fx}^2 - k_x^2)^{3/2} \ldots   \label{moyenne2} \ee
where $k_{Fx}\equiv \sqrt{2m_x\ep_F}$.

\section{St\"uckelberg oscillations}
In the following, we calculate the total inter-band probability associated to the two successive LZ events, in the limit where they can be considered to be separated. This is the case if the system is in the D phase and not too close to the merging. Quantitatively, the Zener tunneling time \cite{revueA}
$\tau \sim \textrm{max}(\sqrt{\delta},\delta)/c_x q_x$  should be shorter than the time $T=2q_D/F=2\sqrt{2m_*|\Delta^*|}/F$ it takes to travel between the two Dirac points, where $\delta=c_x^2 q_x^2/(2c_yF)$ is the adiabaticity parameter.

If the sequence between the two tunneling events is coherent, amplitudes have to be considered instead of probabilities, effectively realizing a coherent St\"uckelberg interferometer \cite{revueA}. In this case, the total probability amplitude to go from the lower to the upper band is the sum of the amplitude for two distinct path. In path 1, the atom jumps to the upper band at the first avoided band crossing (Dirac cone) and then stays in the upper band at the second, such that the amplitude is $A_1=\sqrt{P_Z^y}e^{i\varphi_d}\times e^{i\varphi_1}\times \sqrt{1-P_Z^y}$ where $\sqrt{P_Z^y}e^{i\varphi_d}$ is the amplitude to jump at a single avoided band crossing -- the associated probability of a single LZ event being $P_Z^y$ -- and $\varphi_1=\int_0^T dt E_\textrm{upper band}(t)$ is the phase dynamically acquired by the atom traveling in the upper band between the two Dirac cones. The phase delay $\varphi_d$ which is accumulated at each tunneling event is given by
$\varphi_{d}= -\pi/4 + \delta (\ln \delta -1) + \textrm{arg} \Gamma(1 - i \delta)$ where $\Gamma(z)$ is the gamma function  \cite{revueA}. Up to a $\pi/2$ shift, this phase is the so-called Stokes phase. In path 2, the atom stays in the lower band at the first Dirac cone and then jumps to the upper band at the second. The associated amplitude is  $A_2=\sqrt{1-P_Z^y}\times e^{i\varphi_2}\times \sqrt{P_Z^y}e^{-i\varphi_d}$ where $\varphi_2=\int_0^T dt E_\textrm{lower band}(t)$ is the dynamically acquired phase of the atom traveling in the lower band from one Dirac cone to the other. The total probability is therefore
 \be P_t^y =|A_1+A_2|^2= 4 P_Z^y ( 1 - P_Z^y) \cos^2 (\varphi /2 + \varphi_{d}) \label{LZS}  \ee
where $\varphi=\varphi_1-\varphi_2$ is the dynamically accumulated phase between the two tunneling events
\be \varphi = \int_{0}^T \Delta E(t) d t = {1  \over F} \int_{-q_D}^{q_D} \Delta E(q_y) d q_y \ee
with
$\Delta E\equiv E_\textrm{upper band}-E_\textrm{lower band}$. For the universal Hamiltonian, we have
\be \varphi = {4 \over F} \int_{0}^{\sqrt{2 m^* |\Delta_*|}} d q_y \sqrt{ \left( {q_y^2 \over 2 m^*}+ \Delta_*\right)^2 + c_x^2 q_x^2}  \ee
which can be written in the form
\be \varphi = 4 \sqrt{2 m^*} {|\Delta_*|^{3/2} \over F} I\left( {c_x q_x \over |\Delta_*|} \right) \ee
where the function $I(x)$ is given by $I(x) = \int_0^1 du \sqrt{ (u^2 - 1)^2 + x^2}$ and well approximated by $I(x)\approx \sqrt{4/9+x^2}$. From Eq. (A.9), the incoherent limit (Eq. (6) in the main text) is easily recovered by averaging over the phase.


\begin{thebibliography}{99}
\bibitem{trappingfrequency}Trapping frequencies were measured from dipole oscillations in the optical lattice \cite{Esslinger}. Therefore the band masses replace the bare masses in the harmonic potential energy.
\end{thebibliography}
\begin{thebibliography}{99}
\bibitem[A2]{trappingfrequencyA}The trapping frequencies were measured from dipole oscillations in presence of the optical lattice, therefore it is the band masses that appear in the harmonic potential energy and not the bare atomic masses.
\end{thebibliography}
\end{document}